\documentclass[twocolumn,superscriptaddress,showpacs,pre,floatfix]{revtex4}

\usepackage{graphicx}
\usepackage{amssymb}
\usepackage{amsmath}
\newcommand{\vect}{\vec}
\newcommand{\dd}{\mathrm{d}}
\newcommand{\ee}{\mathrm{e}}
\newcommand{\ii}{\mathrm{i}}
\newcommand{\gb}[3]{J_{#1}(#2,#3)}
\newcommand{\gbn}[1]{J_{#1}}
\newcommand{\gbnxy}{J_{n}(x,y)}
\newcommand{\gbnxysquare}{J_{n}^2(x,y)}
\newcommand{\nmin}{n_{\textrm{min}}}
\newcommand{\nmax}{n_{\textrm{max}}}
\newcommand{\argx}{x}
\newcommand{\argy}{y}
\def\tfrac#1#2{ {\textstyle{\frac{#1}{#2}} } }

\begin{document}

\title{Recursive algorithm for arrays of
generalized Bessel functions:\\
Numerical access to Dirac-Volkov solutions}
\author{Erik L\"{o}tstedt}
\email{Erik.Loetstedt@mpi-hd.mpg.de}
\affiliation{Max-Planck-Institut f\"{u}r Kernphysik, 
Postfach 10 39 80, 69029 Heidelberg, Germany}
\author{Ulrich D. Jentschura}
\affiliation{Max-Planck-Institut f\"{u}r Kernphysik, 
Postfach 10 39 80, 69029 Heidelberg, Germany}
\affiliation{Department of Physics, 
Missouri University of Science and Technology,
Rolla MO65409, USA}

\begin{abstract}
In the relativistic and the nonrelativistic theoretical treatment of 
moderate and high-power laser-matter interaction, the
generalized Bessel function occurs naturally when a
Schr\"{o}dinger-Volkov and Dirac-Volkov solution
is expanded into plane waves. 
For the evaluation of cross sections of quantum electrodynamic
processes in a linearly polarized laser field, 
it is often necessary to evaluate large arrays of generalized Bessel
functions, of arbitrary index but with fixed arguments. 
We show that the generalized Bessel function
can be evaluated, in a numerically stable way, by utilizing a recurrence
relation and a normalization condition only, without having to compute any
initial value. We demonstrate the utility of the method by 
illustrating the quantum-classical correspondence 
of the Dirac-Volkov solutions 
via numerical calculations.
\end{abstract}
 
 
\maketitle
%
%
%
\section{Introduction}
The Volkov solution \cite{Vo1935} is the exact solution of the Dirac equation
in the presence of a classical plane-wave laser field of arbitrary
polarization. In order to evaluate cross sections by quantum electrodynamic
perturbation theory, it is crucial to decompose the Volkov solutions into
plane waves, in order to be able to do the time and space integrations over
the whole Minkowski space-time. If the laser field is linearly polarized, one
naturally encounters the generalized Bessel functions as coefficients in the
plane-wave (Fourier) decomposition of the wave function, both for the
Dirac-Volkov equation as well as for the laser-dressed Klein--Gordon
solutions, and even for Schr\"{o}dinger-Volkov states (see
also Sec.~\ref{Illustrative_Considerations_for_the_Dirac--Volkov_Solutions}). 

The wide use of the generalized Bessel function in theoretical laser physics
is thus due to the fact that different physical quantities, such as
scattering cross sections and electron-positron pair production rates, can be
expressed analytically in terms of infinite sums over generalized Bessel
functions which we here denote by the symbol $\gbnxy$.  The
generalized Bessel function $\gbnxy$ is a generalization of the ordinary Bessel
function $J_n(x)$ and characteristic of the interaction of matter with a
linearly polarized laser field; it depends on two arguments $\argx$ and
$\argy$, and one index $n$. Here, we use it in the convention
\begin{equation} 
\label{def_Jnxy} 
\begin{split} 
\gbnxy = \frac{1}{2\pi}
\int_{-\pi}^\pi \exp[-\ii \, n \, \theta &+ \ii \, \argx \, \sin(\theta)\\ &
- \ii\, \argy \, \sin(2\theta)] \, \dd \theta, 
\end{split} 
\end{equation}
where $n$ is an integer, and $\argx$ and $\argy$ are real numbers. $\gbnxy$
is real valued.  The generalized Bessel functions provide a Fourier
decomposition for expressions of the form $\exp\left[\ii \, \argx \,
\sin\theta - \ii \, \argy \, \sin(2\theta) \right]$ as follows,
\begin{equation} 
\label{fourier} 
\exp\left[\ii \, \argx \, \sin\theta - \ii
\, \argy \, \sin(2\theta) \right] = 
\sum_{n = - \infty}^{\infty} \, \gbnxy \,
\exp\left(\ii \, n \, \theta\right)\,.  
\end{equation}
In practical applications, the angle $\theta$ often has the physical
interpretation of a phase of a laser wave, $\theta = \omega\, t - \vec k
\cdot \vec x$, where $\omega$ is the angular laser photon frequency, and
$\vec k$ is the laser wave vector.  By contrast, the well-known ordinary
Bessel functions are defined as
\begin{equation} 
\label{def_Jnx} 
\gbn{n}(\argx) = \frac{1}{2\pi}
\int_{-\pi}^\pi \exp\left[-\ii \, n \, \theta + 
\ii \, \argx \, \sin(\theta)
\right] \, \dd \theta \,, 
\end{equation}
and they have the fundamental property
\begin{equation} 
\label{fourierJ} 
\exp\left(\ii\,\argx\,\sin\theta\right) =
\sum_{n = - \infty}^{\infty} \, J_n(\argx) \, \exp\left(\ii \, n \,
\theta\right)\,.  
\end{equation}
The generalized Bessel function was first introduced by Reiss in the context
of electron-positron pair creation~\cite{Re1962}, followed by work of
Nikishov and Ritus~\cite{NiRi1964}, and Brown and Kibble~\cite{BrKi1964}. 
Further examples of work utilizing $\gbnxy$ in the relativistic domain
include pair production by a Coulomb field and a laser
field~\cite{Mi1987,MuVoGr2004,SiKrKaEh2006}, laser-assisted
bremsstrahlung~\cite{LoJeKe2007,SchLoJeKe2007,Ro1985}, muon-antimuon
creation~\cite{MuHaKe2008,MuHaKe2008_2}, undulator radiation~\cite{DaVo1993}, and
scattering problems, both classical~\cite{SaSc1970}, and quantum
mechanical~\cite{PaKaEh2002_2,PaKaEh2002}.  A fast and reliable numerical
evaluation of $\gbnxy$  would also speed up calculation of wave packet
evolution in laser fields \cite{RoRoRe2000,RoRoPl2003}. In nonrelativistic
calculations, the generalized Bessel function has been employed mainly for
strong-field ionization
\cite{Re1980,ReKr2003,VanneSaenz2007,GuoChenLiuetal2008}, but also for
high-harmonic generation~\cite{GaSheEd1998,Gaoetal2000}.

On the mathematical side, a thorough study of $\gbnxy$ has been initiated in
a series of papers \cite{DaToLo1990,DaToLo1991,DaChiLoetal1993}, and even
further generalizations of the Bessel function to multiple arguments and
indices have been considered
\cite{DaMaChietal1995,DaToLoetal1998,KoKlWi2006,KoKlWi2007}. On the numerical
side, relatively little work has been performed. Asymptotic approximations
have been found for specific regimes \cite{NiRi1964,Re1980}, and a uniform
asymptotic expansion of $\gbnxy$ for large arguments by saddle-point
integration is developed in Ref.~\cite{Leu1981}.  For some of the
applications described above, in particular when evaluating second-order
laser-assisted quantum electrodynamic processes~\cite{LoJeKe2009}, 
a crucial requirement is to
evaluate large sets of generalized Bessel functions, at fixed arguments
$\argx$ and $\argy$, for all indices $n$ for which the generalized Bessel
functions acquire values which are numerically different from zero (as we
shall see, for $|n| \gg |\argx|, |\argy|$, the generalized Bessel functions
decay exponentially with $n$).

It is clear that recursions in the index $n$ would greatly help
in evaluating large sets of Bessel functions.
For ordinary Bessel functions, an efficient recursive numerical algorithm 
is known, and it is commonly referred to as
Miller's algorithm~\cite{Mi1952,Ga1967}.
However, a generalization of this algorithm for generalized
Bessel functions has been lacking. The purpose of this
paper is to provide such a recursive numerical algorithm: 
We show, using ideas from
\cite{Oliv1968,Matt1980,Matt1982,Wi1984}, that a stable 
recurrence algorithm can indeed
be established, despite the 
more complex recurrence relation satisfied by $\gbnxy$,
as compared to the ordinary Bessel function $J_n(x)$.
The reduction of five-term recursions to four- and three-term 
recursions proves to be crucial in establishing a 
numerically stable scheme.

The computational problem we consider is the following:
to evaluate
\begin{align}
& \gbnxy: \quad \argx \,\, \mbox{fixed}, \quad \argy \,\, \mbox{fixed}\,,
\nonumber\\[2ex]
&\mbox{where} \quad \nmin \le n \le \nmax\,,
\end{align}
by recursion in $n$. Our approach is numerically stable,
and while all algorithms described here have been 
implemented in quadruple precision (roughly 32 decimals),
we note that the numerical 
accuracy of our approach can easily be increased at a small 
computational cost.

Our paper is organized as follows.  In Sec.~\ref{basicprop}, we recall some
well-known basic properties of $\gbnxy$, together with some properties of the
solutions complementary to $\gbnxy$, which fulfill the same recursion relations
(in $n$) as the generalized Bessel functions but have a different asymptotic
behavior for large $|n|$ as compared to $\gbnxy$.  After a review of the Miller
algorithm for the ordinary Bessel function, we present a recursive Miller-type
algorithm for generalized Bessel functions in
Sec.~\ref{Description_of_the_algorithm}, and show that it is numerically
stable. In Sec.~\ref{Results_and_discussion}, we numerically study the accuracy
which can be obtained, and compare the method presented here with other
available methods.  We also complement the discussion by considering in
Sec.~\ref{Illustrative_Considerations_for_the_Dirac--Volkov_Solutions}
illustrative applications of the numerical algorithm for Dirac--Volkov
solutions in particular parameter regions, together with a physical derivation
of the recurrence relation satisfied by the generalized Bessel function.
Section~\ref{Conclusions_and_outlook} is reserved for the conclusions.

%
%

\section{Basic properties of the generalized Bessel function}
\label{basicprop}

%
%
\subsection{Orientation}

Because the definition~\eqref{def_Jnxy} 
provides us with a convenient integral representation 
of the generalized Bessel function, all properties 
of $\gbnxy$ needed for the 
following sections of this article 
can in principle be derived from this representation 
alone~\cite{NiRi1964,Re1980}. E.g., shifting $\theta\to -\theta-\pi$ and 
$\theta\to \theta +\pi$, respectively, in \eqref{def_Jnxy} gives two 
symmetries,
\begin{equation}
\label{symmetries}
\begin{split}
\gb{n}{\argx}{-\argy}&=(-1)^n\gb{-n}{\argx}{\argy},\\
\gb{n}{-\argx}{\argy}&=(-1)^n\gb{n}{\argx}{\argy},
\end{split}
\end{equation}
from which $\gb{-n}{\argx}{\argy}=\gb{n}{-\argx}{-\argy}$ follows. 
We recall the corresponding properties of the ordinary 
Bessel function,
\begin{equation}
\label{symmetriesJ}
J_n(\argx) = (-1)^n J_n(-x) = (-1)^n J_{-n}(x)\,.
\end{equation}
Due to the
symmetries \eqref{symmetries}, we can consider in the following only 
the case of positive $\argx$ and $\argy$ without loss of generality,
provided we allow  $n$ to take arbitrary positive and negative 
integer values. 
Our sign convention for the $y\sin2\theta$-term in the argument of the
exponential in Eq.~\eqref{def_Jnxy} agrees with~\cite{NiRi1964}, but
differs from the one used in~\cite{Re1980}. 
As is evident from inspection of Eqs.~\eqref{def_Jnxy} 
and~\eqref{def_Jnx}, $\gbnxy$ can be expressed as an 
ordinary Bessel function if one of its arguments vanishes,
\begin{equation}
\label{gbnxy_expressed_through_the_usual_Bessel_function}
\gb{n}{\argx}{0}=J_n(\argx),
\qquad\gb{n}{0}{\argy}=
\left\{\begin{array}{ll}
J_{-n/2}(\argy)&\textrm{if }n\textrm{ even}\\
0&\textrm{if }n\textrm{ odd}.
\end{array}\right.
\end{equation}
By inserting the expansion of the ordinary Bessel function 
$\sum_{n = - \infty}^\infty 
J_n(\argx)\exp(in\theta)=\exp(\ii \, \argx \, \sin\theta)$ into
Eq.~\eqref{def_Jnxy}, we see that $\gbnxy$ can be expressed as an
infinite sum of products of ordinary Bessel function,
\begin{equation}
\label{gen_bes_as_sum_usual_bes}
\gbnxy=\sum_{s=-\infty}^\infty J_{2s+n}(\argx)J_s(\argy).
\end{equation}
There are also the following sum rules,
\begin{equation}
\label{sumruleJnxy}
\sum_{n=-\infty}^\infty \gbnxy =
\sum_{n=-\infty}^\infty \gbnxysquare = 1\,,
\end{equation}
which can be derived by considering the case
$\theta = 0$ in Eq.~\eqref{fourier} 
[for $\sum_{n=-\infty}^\infty \gbnxy = 1$],
and by considering Eq.~\eqref{fourier} multiplied with  its complex
conjugate, and integrating over one 
period [for $\sum_{n=-\infty}^\infty \gbnxysquare = 1$].
The relation~\eqref{sumruleJnxy} is important 
for a recursive algorithm, because it provides a normalization 
for an array of generalized Bessel function 
computed according to the recurrence relation
\begin{equation}
\label{recrel}
\begin{split}
& 2 n \gbnxy = \argx \,
\left[ \gb{n+1}{\argx}{\argy} + 
\gb{n-1}{\argx}{\argy}] \right. \\
& \qquad \left. -2\argy \, [\gb{n+2}{\argx}{\argy} +
\gb{n-2}{\argx}{\argy}
\right]\,,
\end{split}
\end{equation}
which connects generalized Bessel functions of the same arguments
but different index $n$.
Equation \eqref{recrel} can be derived by partial integration of
Eq.~\eqref{def_Jnxy}. Interestingly, Eq.~\eqref{recrel} together with the
normalization condition~\eqref{sumruleJnxy} can be taken as an alternative
definition for $\gbnxy$, from which the integral representation~\eqref{def_Jnxy}
follows. The recursion \eqref{recrel}
is the basis for the algorithm described below
in Sec.~\ref{Description_of_the_algorithm}.

%
%

\subsection{Saddle point considerations}
\label{Saddle_point_considerations}

A qualitative picture of the behavior of $\gbnxy$ as a function of $n$ can be
obtained by considering the position of the saddle points of the integrand in
\eqref{def_Jnxy}~\cite{Leu1981,KoKlWi2006}. By definition, a saddle point
$\theta_s$ denotes the point where the derivative of the argument of the
exponential in~\eqref{def_Jnxy} vanishes, and therefore satisfies
\begin{equation}
\label{saddle_point_equation} 
\cos \theta_{s\pm} =
\frac{\argx}{8\argy}\pm\sqrt{\frac{\argx^2}{64 \, \argy^2}
+\frac{1}{2}-\frac{n}{4\argy}}.  
\end{equation}
By writing $\gbnxy$ as
\begin{equation}
\label{def2_Jnxy}
\begin{split}
\gbnxy = \frac{1}{\pi}\mathrm{Re}\Bigg\{ \int_{0}^\pi 
\exp[-\ii \, n \, \theta &+ \ii \, \argx \, \sin(\theta)\\
& - \ii\, \argy \, \sin(2\theta)] \, \dd \theta\Bigg\},
\end{split}
\end{equation}
we can consider only saddle points with $0\le \mathrm{Re}\,\theta_s \le \pi$.
By the properties of the cosine function, the saddle points come in conjugate
pairs, so that if $\theta_s$ is a saddle point, so is $\theta_s^\ast$.
Furthermore, since $\cos(2\pi-\theta_s)=\cos\theta_s$, the saddle points are
placed mirror symmetrically around $\mathrm{Re}\,\theta_s=\pi$.  Since each of
the endpoint contributions at $\theta=0$ and $\theta=\pi$ to the integral
\eqref{def2_Jnxy} vanish (provided the endpoints are not saddle points), an
asymptotic approximation for $\gbnxy$ is provided by the saddle point method
(the method of steepest descent)~\cite{Olv1997}, by summing the contributions
from the saddle points $\theta_{s}$ situated on the path of steepest descent.
Here, imaginary saddle points (i.e., saddle points with ${\rm Im} \,
\theta_{s\pm} \neq 0$) give exponentially small contributions to the integral,
while real saddle points contribute with an oscillating term. Closer inspection
of Eq.~\eqref{saddle_point_equation} reveals two cases.

\begin{figure*}[thb]
\begin{center}
\includegraphics[width=0.45\linewidth]{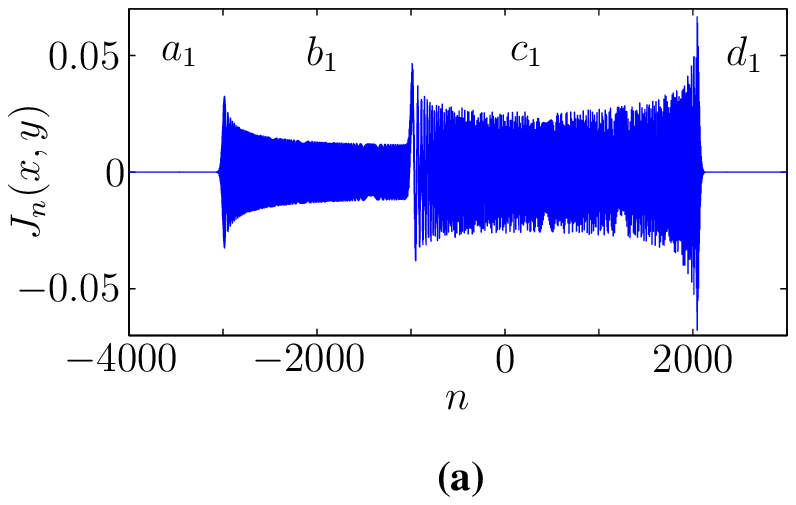}
\hspace{1cm}
\includegraphics[width=0.45\linewidth]{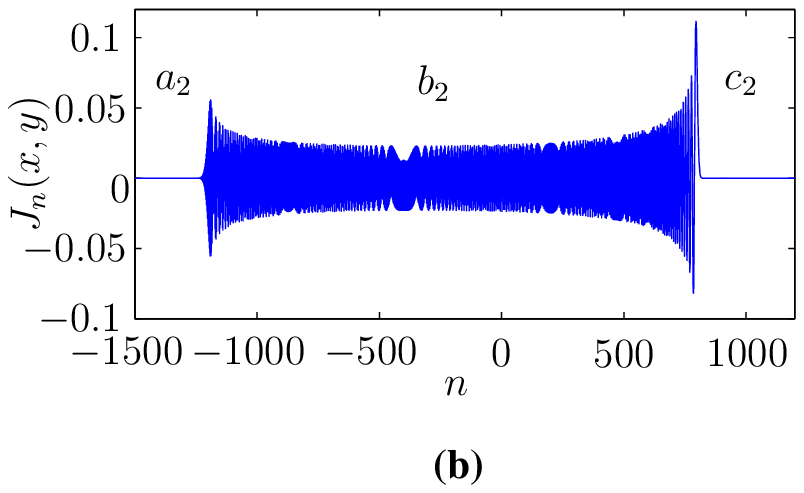}
\\[2ex]
\caption{\label{fig1}(Color online.) Illustration of the different 
saddle point regions of $\gbnxy$, for the
two qualitatively different cases number 1,
with $8 y > x$ (here $x=y=10^3$ was used), and case number 2, with $8 y < x$ 
($x=10y=10^3$).
In case 1 [panel (a)],
the transition from region $b_1$, where only one saddle point is
real, to $c_1$, where two real saddle points contribute, occurs 
precisely at $n=-2\argy+\argx=-10^3$
(see also Table~\ref{table1}). 
The complex oscillating behavior in region $c_1$
can be understood as interference between the contributions from the two 
real
saddle points. In case 2 [panel (b)], we have $8\argy<\argx$, with
only three as opposed to four 
qualitatively different regions (see also Table~\ref{table1}).}
\end{center}
\end{figure*}

In case 1, with $8\argy>\argx$, there are four different regions, which we
denote by $a_1, b_1, c_1, d_1$ (see Table~\ref{table1}). In region $a_1$, where
$n<-2\argy-\argx$, we have four distinct saddle points solutions
$\theta_{s\pm}$, $\theta_{s\pm}^\ast$, which are all imaginary, and 
$\gbnxy$ is
exponentially small. Region $b_1$, where $-2\argy -\argx <n<-2\argy +\argx$,
has two imaginary ($\theta_{s+}$, $\theta_{s+}^\ast$) and one real saddle point
$\theta_{s-}$, and $\gbnxy$ exhibits an
oscillating behavior here. For $-2\argy +\argx <n< 2\argy +\argx^2/(16\argy)$,
i.e.~in region $c_1$, both saddle points are real, and in the region $d_1$,
$n>2\argy +\argx^2/(16\argy)$, the two saddle points $\theta_{s\pm}$ are again imaginary, which
results in very small numerical values of the generalized Bessel functions.
For case 2, $8\argy<\argx$, there are only three regions, as recorded in
Table~\ref{table1}.  The two cases coincide if $8\argy=\argx$.
Figure~\ref{fig1} illustrates the two different cases.

\begin{center}
\begin{table}
\begin{center}
\caption{\label{table1}
Saddle-point configurations for the generalized Bessel function
$\gbnxy$ as a function of the arguments $\argx$ and $\argy$.
A distinct imaginary saddle point is denoted ``imag.'' whereas a real
saddle point is denoted ``real.'' The different regions are 
illustrated in Fig.~\ref{fig1}.}
\begin{tabular}{c@{\hspace{0.4cm}}c@{\hspace{0.4cm}}c}
\hline
\hline
\multicolumn{3}{l}{
\rule[-3mm]{0mm}{8mm}
Case 1: $8y > x$} \\
\hline
region & condition & saddle points \\
\hline
\rule[-2mm]{0mm}{6mm}
$a_1$ & $n<-2\argy-\argx$ 
& 4 imag. \\
\rule[-2mm]{0mm}{6mm}
$b_1$ & $-2\argy -\argx <n<-2\argy +\argx$ 
& 2 imag.$+$real\\
\rule[-2mm]{0mm}{6mm}
$c_1$ & $-2\argy +\argx <n< 2\argy +\argx^2/(16\argy)$
& 2 real\\
\rule[-2mm]{0mm}{6mm}
$d_1$ & $n>2\argy +\argx^2/(16\argy)$ 
& 2 imag.\\
\hline
\hline
\multicolumn{3}{l}{
\rule[-3mm]{0mm}{8mm}
Case 2: $8y < x$} \\
\hline
region & condition & saddle points \\
\hline
\rule[-2mm]{0mm}{6mm}
$a_2$ & $n<-2\argy-\argx$ 
& 4 imag. \\
\rule[-2mm]{0mm}{6mm}
$b_2$ & $-2\argy -\argx <n<-2\argy +\argx$ 
& 2 imag.$+$real\\
\rule[-2mm]{0mm}{6mm}
$c_2$ & $n> -2\argy +\argx$
& 2 imag.\\
\hline
\hline
\end{tabular}
\end{center}
\end{table}
\end{center}

%
%

In all regions $a_1, b_1, c_1, d_1, a_2, b_2, c_2$, 
there are, depending on the region, up to four saddle points to consider. 
Of these one or two saddle points contribute to the numerical approximation to 
$\gbnxy$.
For large arguments $\argy$, $\argx$ and/or a large index
$n$, asymptotic expressions 
can be derived~\cite{Leu1981,KoKlWi2006}. The
general form for the leading-order term is (see  \cite{Olv1997} for a clear 
exposition of the general theory of asymptotic expansions of special functions)
\begin{equation}
\label{asymptotic_approximation}
\begin{split}
& \gbnxy \\
& \approx
\mathrm{Re} \left[
\sqrt{\frac{2}{\pi|f''(\theta_{s+})|}} \,
\exp\left( \ii f(\theta_{s+})-\ii \, n \, \theta_{s+} + 
\ii \varepsilon_+ \right)
\right. \\
& \quad \left. +
\sqrt{\frac{2}{\pi|f''(\theta_{s-})|}} \,
\exp\left( \ii f(\theta_{s-})-\ii \, n \, \theta_{s-} +
 \ii  \varepsilon_- \right)
\right] \\[2ex]
& = F_+(n,x,y)+F_-(n,x,y),
\end{split}
\end{equation}
where 
\begin{equation}
f(\theta) = \argx\sin(\theta) - \argy\sin( 2\theta )\,. 
\end{equation}
For imaginary 
saddle points, only the contribution of those situated on the path of 
steepest descent should be included 
, i.e., the 
integration around the 
saddle point should be carried out along a curve of constant 
complex phase, with $\theta=\mu+i\nu$ satisfying
\begin{equation}
\label{path_of_steepest_descent}
\mathrm{Im}\left(\ii f(\theta) - \ii \, n \, \theta\right) = \mathrm{const.}
\end{equation}
on that curve. In practice this means for regions with imaginary
saddle points only, $\gbnxy$ is given by the 
contribution from the saddle point with smallest $|\mathrm{Im}\,\theta_s|$,
and in regions with both imaginary and real $\theta_s$ the contribution 
from the imaginary saddle point can be neglected. However, we will see
in the following discussion that all saddle points, including those not 
on the path of steepest descent which would produce an 
``exponentially large'' contribution, can be interpreted in terms of
complementary solutions to the recurrence relation \eqref{recrel}.
The constant phase $\varepsilon_\pm$ in Eq.~\eqref{asymptotic_approximation} 
is given by
\begin{equation}
\varepsilon_\pm = 
\frac{\pi}{4} \, {\rm sgn} \left[f''(\theta_{s\pm}) \right]
\end{equation}
for real saddle points. For imaginary saddle points, $\varepsilon_\pm$ can be 
found from the requirement
\begin{equation}
\tan \varepsilon_\pm=\left.\frac{d \nu}{d\mu}\right|_{\theta=\theta_{s\pm}},
\end{equation}
with $\theta=\mu+i\nu$ describing the path of steepest descent 
[see Eq.~\eqref{path_of_steepest_descent}]. For a detailed treatment of the 
saddle point approximation of $\gbnxy$ we refer to \cite{Leu1981}, where 
uniform approximations, valid also close to the turning points (the borders 
between the regions described in Fig.~\ref{fig1}), and beyond the leading 
term \eqref{asymptotic_approximation}, are derived. For our purpose, namely 
to identify the asymptotic behavior of the complementary solutions, the 
expression \eqref{asymptotic_approximation} is sufficient.
%
%
\subsection{Complementary solutions}
The recurrence relation \eqref{recrel} involves
the five generalized Bessel functions of indices $n-2,n-1,n,n+1,n+2$.
In general, an $m$-term recursion relation is said to be of 
order $m-1$. If we regard the index $n$ as a continuous variable, 
then a recursion relation of order $m-1$
corresponds to a differential equation of order $m-1$,
which has $(m-1)$ linearly independent solutions.
Equation~\eqref{recrel} 
consequently has four linearly independent 
(complementary) solutions. The function $\gbnxy$ is one of these.

For the analysis of the recursive algorithm in
Sec.~\ref{Description_of_the_algorithm} below, we should also identify the
complementary solutions to the recurrence relation \eqref{recrel}.  For our
purposes, it is sufficient to recognize the asymptotic behavior of the
complementary solutions in the different regions $a_1$--$d_1$ and $a_2$--$c_2$
(see Fig.~\ref{fig1}).  It is helpful to observe that the recurrence relation
\eqref{recrel} is satisfied asymptotically by each term $F_{\pm}(n,x,y)$ from
Eq.~\eqref{asymptotic_approximation} individually. The recurrence relation
\eqref{recrel} is also satisfied, asymptotically, by a function obtained by
taking in Eq.~\eqref{asymptotic_approximation} a saddle point that is not on
the path of steepest descent, which is equivalent to changing the sign of the
entire argument of the exponential. In addition, for real saddle points  and in
regions with only two imaginary saddle points, the recurrence relation is
asymptotically satisfied by taking the same saddle point but the
imaginary part instead of the real part in Eq.~\eqref{asymptotic_approximation}
(and thereby changing the phase). 

In regions with four imaginary
saddle points ($a_1, a_2$), there are 
thus two solutions that are
exponentially increasing with the index $n \to -\infty$
[the two solutions correspond to the two saddle points $\theta_s$ where
$\mathrm{Re}(\ii f(\theta_s) - \ii n \theta_s)>0$], 
and two further solutions which are exponentially decreasing
[From $\theta_s$ with $\mathrm{Re}(\ii f(\theta_s) - \ii n \theta_s)<0$]. 
In regions with two imaginary and one real saddle point ($b_1, b_2$), 
the four solutions behave as follows.
There are two oscillatory solutions 
[these correspond to the real and imaginary parts
of the term which contains the real saddle point in 
Eq.~\eqref{asymptotic_approximation}],
and a third solution which is 
exponentially increasing, and a fourth one which is exponentially
decreasing as $n \to -\infty$ [the two latter 
solutions are due to  the imaginary saddle points in 
Eq.~\eqref{asymptotic_approximation}].
The region with two real saddle points ($c_1$) has four
oscillating solutions, as a function of $n$. 
Finally, in regions $d_1$ and $c_2$, where we have two distinct imaginary
saddle points, we have two exponentially increasing (as $n \to \infty$)
solutions with 
different phase,
and two exponentially decreasing with different phase.
Concerning the question of how to join the different asymptotic behaviors to
form four linearly independent solutions, we note that $\gbnxy$ is the only
solution which can decrease in both directions $n\to \pm \infty$, since it 
represents the unique, normalizable physical solution to the wave equation
(see subsection \ref{Physical_origin_of_the_recrel}). Furthermore, there
must be one solution that increases exponentially where $\gbnxy$ decreases,
and that exhibits an oscillatory behavior where $\gbnxy$ also oscillates. The 
reason is that in either of the limits $x\to 0$ or $y\to 0$, we must recover 
the ordinary Bessel function, and the Neumann function as the two solutions
to the recurrence relation. Having fixed the asymptotic behavior of two
of the solutions, the behavior of the the two remaining functions follows.
We label the four different solutions with 
$\gbn{n}$, $Y_n$, $X_n$, and $Z_n$, where $\gbn{n}$ is the 
generalized Bessel function $\gbnxy$ with the 
arguments $\argx$, $\argy$ suppressed. 

Integral representations for the complementary solutions can be 
found by employing Laplace's method \cite{Jo1960}, details of which
will be described elsewhere. The explicit expressions can be found in
the Appendix.
However, as noted previously, in this paper we shall
need only the asymptotic properties of the complementary solutions, which can
be deduced from \eqref{asymptotic_approximation}. 

We also observe that the situation for $\gbnxy$
described above is directly analogous to that of the ordinary Bessel 
function $J_n(x)$ and the complementary Neumann (also called Weber)
function $Y_n(x)$
of a single argument. For $x\gg n$ they have the asymptotic behavior 
$J_n(x)\approx \mathrm{Re}\,\sqrt{2/(\pi x)}\exp(\ii x-\ii\pi/4-\ii n\pi/2)$, 
$Y_n(x)\approx \mathrm{Im}\,\sqrt{2/(\pi x)}\exp(\ii x-\ii \pi/4-\ii n\pi/2)$,
and for  $x\ll n$ we have $J_n(x)\approx (\ee x)^n (2n)^{-n}/\sqrt{2\pi n}$,
$Y_n(x)\approx -2(\ee x)^{-n} (2n)^{n}/\sqrt{2\pi n}$.
  
\begin{figure}
\begin{center}
\includegraphics[width=0.95\linewidth]{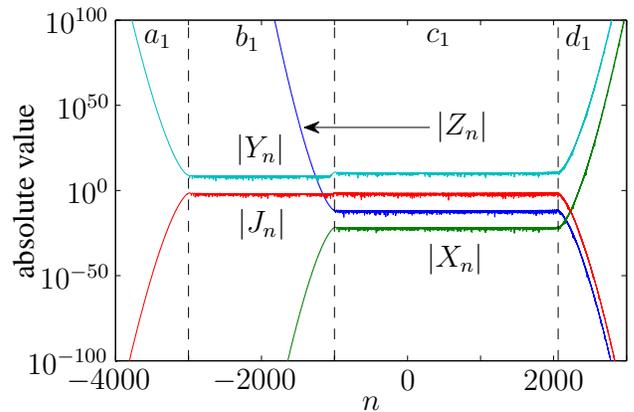}
\caption{\label{fig2}
(Color online.) The five-term recurrence relation \eqref{recrel}
has four linearly independent solutions. Note the logarithmic scale. 
The values $\argx = \argy = 10^3$ were used for the calculation, 
corresponding to case 1 ($8y > x$). The
solutions are labeled by $\gbn{n}$ [red line, the true generalized Bessel
function $\gbnxy$], $Y_n$ (light blue line), $X_n$ (green line), and $Z_n$ (blue
line). The numerically obtained solutions $X_n$, $Y_n$, and $Z_n$ have been
shifted vertically by multiplication with an appropriate constant (of order 
$10^{10}$ for $Y_n$, $10^{-10}$ for $Z_n$, and $10^{-20}$ for $X_n$) for clarity. 
The separation of the different saddle point regions is marked with dashed
lines. The regions $a_1, b_1, c_1$ and $d_1$ 
are described in Table~\ref{table1}.} 
\end{center}
\end{figure}

According to the above discussion and as illustrated 
in Fig.~\ref{fig2}, the functions $\gbn{n}$, $Y_n$, $X_n$, and $Z_n$ have the following relative
amplitudes in the different regions:
\begin{equation}
\label{relative_amplitudes}
\begin{array}{ll}
\textrm{region } a_1, a_2: & |Z_n|>|Y_n|>|\gbn{n}|>|X_n| ,\\
\textrm{region } b_1, b_2: & |Z_n|>|Y_n|\sim|\gbn{n}|>|X_n|,\\
\textrm{region } c_{1}:    & |Z_n|\sim|Y_n|\sim|\gbn{n}|\sim|X_n|,\\
\textrm{region } d_{1},c_2:& |Y_n|\sim|X_n|>|\gbn{n}|\sim|Z_n|.
\end{array}
\end{equation}
In Eq.~\eqref{relative_amplitudes}, we have assumed that all functions have 
the same order of magnitude in the oscillating region. 
This can be accomplished by choosing a suitable constant prefactor
for the complementary functions $Z_n$, $Y_n$, and $X_n$. 
Figure~\ref{fig2} shows an example of the four different solutions,
for case 1 ($8y > x$). 
The actual numerical computation of the complementary solutions
is discussed in Sec.~\ref{Demonstration_of_numerical_stability}.

%
%
\section{Miller--Type algorithm for generalized Bessel functions}
\label{Description_of_the_algorithm}
%
%
\subsection{Recursive Miller's algorithm for ordinary Bessel functions}
\label{Millers_algorithm_for_the_usual_Bessel_function}

A straightforward implementation of  Miller's
algorithm \cite{Mi1952,Ga1967,Mo1974_2}
can be used for the numerical calculation of the
ordinary Bessel function $J_n(\argx)$. We note that there are also other ways
of numerically evaluating $J_n(\argx)$, 
which include series expansions \cite{Wa1962} or contour 
integration \cite{Ma1993}. In the following, we review the simplest 
form of Miller's algorithm, to prepare for the discussion on the generalized 
algorithm. We treat the case of
positive $n$ and $\argx$.
For negative $n$ and $\argx$, we appeal to 
the symmetry relation \eqref{symmetriesJ}. 
The properties of $J_n(\argx)$ used for the algorithm are
the recurrence relation \eqref{recrel}, with $\argy=0$,
which automatically reduces \eqref{recrel} to a three-term relation
with only two linearly independent solutions.
We also use the 
normalization condition $\sum_{n=-\infty}^\infty J_n(\argx)=1$. 

Viewed as a function of $n$,
$J_n(\argx)$ exhibits an oscillatory behavior for $n<\argx$, and decreases
exponentially for $n>\argx$. 
The complementary solution $Y_n(\argx)$, called the
 Neumann function, oscillates for $n<\argx$ and grows exponentially for
$n>\argx$. To calculate an array of $J_n(\argx)$, for $0 \le n < N$, 
with $N > x$, we proceed as follows. We
take a (sufficiently large) integer $M>N$, and the  initial
values $c_M=1$, $c_{M+1}=0$. We use the recurrence relation
\eqref{recrel} with $\argy=0$ to calculate all $c_n$ with
indices $0\le n<M$ by downward recursion in $n$. 
Now, since the ensemble of the 
$Y_n(\argx)$ plus the $J_n(\argx)$ constitute a complete basis set of
functions satisfying the recurrence relation, the computed array 
of the $c_n$ can be
decomposed into a linear combination,
\begin{equation}
\label{lin_comb}
c_n = \alpha \, J_n(\argx) + \beta \, Y_n(\argx)\,,
\end{equation}
where $\alpha$ and $\beta$ are  constants,
and this decomposition is valid for any $n$.
That means that the same decomposition 
must also be valid for the initial index $M + 1$ from which 
we started the downward recursion, i.e.
\begin{equation}
c_{M+1}=0 = \alpha \, J_{M+1}(\argx)+\beta \, Y_{M+1}(\argx).
\end{equation}
From Eq.~\eqref{lin_comb} it follows that
\begin{equation}
\label{solution_Miller_usual_Bessel_function}
c_n = 
\alpha \, \left(J_n(\argx) -
\frac{J_{M+1}(\argx) \, Y_n(\argx)}{Y_{M+1}(\argx)}
\right)\,.
\end{equation}
Provided the starting index $M>x$ is chosen large enough, the quantity
$J_{M+1}(\argx)/Y_{M+1}(\argx)$ is a small quantity, due to the exponential
character of $J_n(\argx)$ and $Y_n(\argx)$ for index $n>x$, so that the computed
array $c_n$ is to a good approximation proportional to the sought $J_n(\argx)$.
Loosely speaking, we can say that we have selected the exponentially
decreasing function 
$J_n(\argx)$ by the downward recursion, because the 
exponentially increasing function $Y_n(\argx)$ as $|n| \to \infty$ 
is suppressed in view of its exponential decrease for 
decreasing $|n|$.
In other words, the error introduced by
the initial conditions decreases rapidly due to the rapid decrease of
$Y_n(\argx)$ for decreasing $n$, so that effectively only the part proportional
to $J_n(\argx)$ is left.

Finally, the constant $\alpha$ can be found by imposing  the normalization
condition 
\begin{equation}
\sum_n c_n=c_0+2\sum_{n=1}^\infty c_{2n}=1.
\end{equation}
Here, we have used the symmetries \eqref{symmetriesJ}
in order to eliminate the terms of odd index from the sum.

Remarkably, numerical values of $J_n(\argx)$ can be computed by
using only the recurrence relation and the normalization condition, 
and not a single initial
value is needed [e.g., one might otherwise imagine 
$J_0(\argx)$ to be calculated by a series expansion].
Miller's algorithm has subsequently been refined and the error propagation
analyzed by several authors \cite{Ga1967,Olv1964,Oliv1967,OlvSoo1972}, and also
implemented \cite{RaFe1993,To1993,YoMe1997}. 

%
%
\subsection{Recursive algorithm for generalized Bessel functions}

In view of the four different solutions pictured in Fig.~\ref{fig2},
it is clear from the discussion in the preceding subsection that $\gbnxy$ 
cannot be calculated by na\"{\i}ve 
application of the recurrence relation. 
The general paradigm (see Fig.~\ref{fig2}) therefore has to change.
We first observe that if 
we would start the recursion using the five-term 
recurrence relation \eqref{recrel} in the downward direction,
starting from large positive $n$,
then the solution would eventually pick up a component proportional
to $Z_n$, which diverges for $n\to -\infty$. Conversely, if 
we would start the recursion using the five-term 
recurrence relation \eqref{recrel} in the upward direction,
starting from large negative $n$,
then the solution would pick up a component proportional
to $X_n$. Thus, Eq.~\eqref{recrel} cannot be used directly.

The solution to this problem is based on rewriting 
\eqref{recrel} in terms of recurrences with less
terms (only three or four as opposed to five).
By consequence, the reformulated recurrence has less 
linearly independent solutions, and in fact it can be shown 
(see the discussion below) that the four-term recurrence,
if used in the appropriate directions in $n$,
numerically eliminates the most problematic solution $Z_n$
which would otherwise be admixed to $\gbnxy$
for $n \to \infty$, 
leading to an algorithm by which it is
possible to calculate the generalized
Bessel function $\gbnxy$ for $n$
down to the point where we transit from region $b_1$ to $a_1$ 
in Fig.~\ref{fig2}, where the recurrence invariably
picks up a component from the exponentially growing solution
$Y_n$, and it becomes unstable.
However, by using the additional three-term 
recurrence in suitable directions in $n$, 
we can numerically eliminate 
the remaining problematic solution $Y_n$
which would otherwise be admixed to $\gbnxy$ for $n \to -\infty$
even after the elimination of $Z_n$, 
leading to an algorithm by which it is
possible to calculate the generalized
Bessel function $\gbnxy$ for  $n$
up to the point where we transit from region $c_1$ to $d_1$ 
in Fig.~\ref{fig2}, where the recurrence invariably
picks up a component from the exponentially growing solution
$X_n$, and it becomes unstable.
In the end, we match the results
of the four-term recursion and the three-term recursion at some 
``matching index'' $K$, situated in region $b_1$ or $c_1$, normalize 
the solutions 
according to Eq.~(\ref{sumruleJnxy}), and obtain numerical
values for $\gbnxy$.   

Indeed, in region
$b_1$ (see Fig.~\ref{fig2}), the wanted solution $\gbnxy$ satisfies
$|X_n/X_{n+1}|<|\gbnxy/\gb{n+1}{\argx}{\argy}|<|Z_n/Z_{n+1}|$, which 
means that
here application of the recurrence relation is unstable in both the 
upward and downward directions with respect to $n$. 
By a suitable transformations of the recurrence relation, we 
remove one, and then
two of the unwanted solutions $Y_n$ and $Z_n$. With only three (or two)
solutions left, we can proceed exactly as described in subsection
\ref{Millers_algorithm_for_the_usual_Bessel_function} to calculate $\gbnxy$ 
in a
stable way by downward 
recursion in $n$. We note that the general case of stable
numerical solution of recurrence relations of arbitrary order has been 
described
previously in \cite{Oliv1968,Matt1980,Matt1982,Wi1984,Oliv1968_2}, but the 
application
of this method to the calculation of the generalized Bessel function has not
been attempted before, to the authors knowledge.

In the following, we describe the algorithm to compute an approximation to the
array $\gb{n}{\argx}{\argy}$, $\nmin\le n \le \nmax$. We let 
\begin{equation}\label{nminus_nplus}
n_- = - 2\argy - \argx, \qquad n_+ =
\left\{\begin{array}{ll}
2\,\argy + {\displaystyle \frac{\argx^2}{16\,\argy}}
                  & \textrm{if} \;\; 8\argy > x \\[2ex]
-2\,\argy + \argx & \textrm{if} \;\; 8\argy < x
\end{array}\right.
\end{equation}
denote the ``cutoff'' indices, beyond which $\gbnxy$ decreases exponentially 
in
magnitude. In terms of the regions introduced in Table~\ref{table1}, $n_-$ 
marks the transition from region $a_1$ to $b_1$ for case 1 (or $a_2$ to $b_2$ 
for case 2), and $n_+$ is the border between region $c_1$ and $d_1$ for case 1 
(between $b_2$ and $c_2$ in case 2). Note that we do not assume $\nmin < n_-$ 
or $\nmax> n_+$, in general $\nmin$ and $\nmax$ are arbitrary 
(with $\nmin\le\nmax$). The usual situation is however to require $\nmin \le n_-$ 
and $\nmax\ge n_+$. 
Without loss of generality, we assume that both $\argx$
and $\argy$ are nonzero [otherwise the problem
reduces to the calculation of ordinary Bessel functions via
Eq.~\eqref{gbnxy_expressed_through_the_usual_Bessel_function}]. 

Central for our algorithm is the transformation of the
five-term recurrence relation \eqref{recrel} into a four-term and
three-term recurrence relation. Suppressing the 
arguments $\argx$ and $\argy$, we can write
the four-term recurrence
\begin{equation}
\label{four-term_backwards}
2 \, \argy \, \gbn{n+1}+\xi^1_n \, \gbn{n}+\xi^2_n \, \gbn{n-1}+
\xi^3_n\, \gbn{n-2}=0,
\end{equation}
and the second-order relation
\begin{equation}\label{three-term_backwards}
2\,\argy \,\gbn{n+1} +\lambda^1_n \, \gbn{n} +\lambda^2_n \, \gbn{n-1}=0\,.
\end{equation}
The coefficients themselves also satisfy recursion relations,
which are however of first order, namely
\begin{equation}
\label{zeta_eta_xi_backwards}
\begin{split}
\xi^1_n &= -\argx-\frac{4\,\argy^2}{\xi^3_{n-1}},
\qquad 
\xi^2_n=2(n-1)-\frac{2\,\argy\,\xi^1_{n-1}}{\xi^3_{n-1}},\\
\xi^3_n &= -\argx-\frac{2 \, \argy \, \xi^2_{n-1}}{\xi^3_{n-1}},
\end{split}
\end{equation}
and
\begin{equation}
\label{kappa_lambda_backwards}
\lambda^1_n = \xi^1_n - \frac{2 \, \argy \, \xi^3_n}{\lambda^2_{n-1}},
\qquad
\lambda^2_n=\xi^2_n-\frac{\lambda^1_{n-1} \, \xi^3_n }{\lambda^2_{n-1}}\,.
\end{equation}

By construction, all sequences $y_n$ that solve 
the original recurrence relation \eqref{recrel}, 
also solve Eq.~\eqref{four-term_backwards}
and Eq.~\eqref{three-term_backwards}, 
regardless of the initial conditions used to
calculate the coefficients $\xi^{1,2,3}_n$ and  $\lambda^{1,2}_n$.
The converse does
not hold: a solution $y_n$ to the transformed recurrence relation
\eqref{four-term_backwards} or \eqref{three-term_backwards} 
does not automatically solve Eq.~\eqref{recrel}. Rather, this 
depends on the initial conditions used for the
coefficients $\xi^{1,2,3}_n$ (or $\lambda^{1,2}_n$).

The transformation into a four-term and three-term relation offers
a big advantage, as briefly anticipated above. 
We now describe how the algorithm is
implemented in practice, and postpone the discussion of numerical stability
until subsection \ref{Demonstration_of_numerical_stability}.
We proceed in five steps.
\begin{itemize}
\item[1.] Select a positive starting index $M_+ > n_+,\nmax$ and a negative
starting index $M_- < n_-,\nmin$, where the $M$'s differ from the 
$n$'s by some ``safety margin.'' The dependence of the accuracy obtained on
the ``safety margin'' is discussed later, in 
Sec.~\ref{Results_and_discussion}.
\item[2.] Calculate the arrays $\xi^{1,2,3}_n$, and $\lambda^{1,2}_n$ for
$M_-\le n\le M_++1$, employing the recurrence relations
\eqref{zeta_eta_xi_backwards} and  \eqref{kappa_lambda_backwards} in the
upward direction of for $n$. The recurrence is started at $n=M_-$ with
nonzero $\xi_{M_-}^3$, but otherwise arbitrary initial values. 
A practically useful choice is 
$\xi^1_{M_-}=\xi^2_{M_-}=\xi^3_{M_-}=1$ for the 
four-term formula \eqref{zeta_eta_xi_backwards} and 
$\lambda^1_{M_-}=\lambda^2_{M_-}=1$
for the three-term recurrence \eqref{kappa_lambda_backwards}.
\item[3.] Calculate the array $f_n$, with $M_-\le n\le M_+$ according to the
recurrence formula \eqref{four-term_backwards}, and the array 
$g_n$, also with $M_-\le n\le M_+$ according 
to the recurrence formula \eqref{three-term_backwards}. In
both cases, the recurrence is performed in the downward directions for 
$n$, with arbitrary (but not all zero) starting values
for $f_{M_+}$, $f_{M_++1}$, $f_{M_++2}$ and $g_{M_+}$, $g_{M_++1}$.
\item[4.] Choose a ``matching index'' $K$, 
with $n_- < K < n_+$ to match the solutions 
$g_n$ and $f_n$ to each other, realizing that 
$g_n$ will be unstable for $n >n_+$, and $f_n$ will be unstable for $n <n_-$. 
Specifically, we construct the array
\begin{equation}
h_n = \left\{
\begin{array}{ll}
g_n & \textrm{if } \quad M_- \le n \le K \,, \\[2ex]
{\displaystyle \frac{g_K}{f_K} \, f_n} &
      \textrm{if } \quad K < n \le M_+\,,
\end{array}\right.
\end{equation}
where $f_K,g_K\neq 0$ is assumed.
\item[5.] The numerical approximation to the generalized Bessel functions is 
now
given by normalizing $h_n$ according to the sum rule \eqref{sumruleJnxy},
\begin{equation}
\begin{split}
& \gbnxy \approx 
\mathrm{sgn} \left(\frac{h_n}{H_1} \right)\, \sqrt{\frac{h_n^2}{H_2}},\\
& \nmin\le n\le\nmax,\qquad H_j = \sum_{n=M_-}^{M_+} h_n^j.
\end{split}
\end{equation}
The reason why we normalize the sum of squares is that a summation of
only nonnegative terms cannot suffer from numerical cancellation. 
An alternative way of normalization would consist in
calculating a particular value of $\gbnxy$, say
$\gb{0}{\argx}{\argy}$, by another method, like the sum
\eqref{gen_bes_as_sum_usual_bes}, or an asymptotic expansion \cite{Leu1981}.
In this case, the approximation to $\gbnxy$ would be given as
\begin{equation}
\gbnxy \approx \frac{\gb{0}{\argx}{\argy}}{h_0} h_n \,,
\end{equation}
for all $\nmin\le n\le\nmax$,
provided $\gb{0}{\argx}{\argy},h_0\neq 0$.
\end{itemize}
\begin{figure*}
\begin{center}
\includegraphics[width=0.45\linewidth]{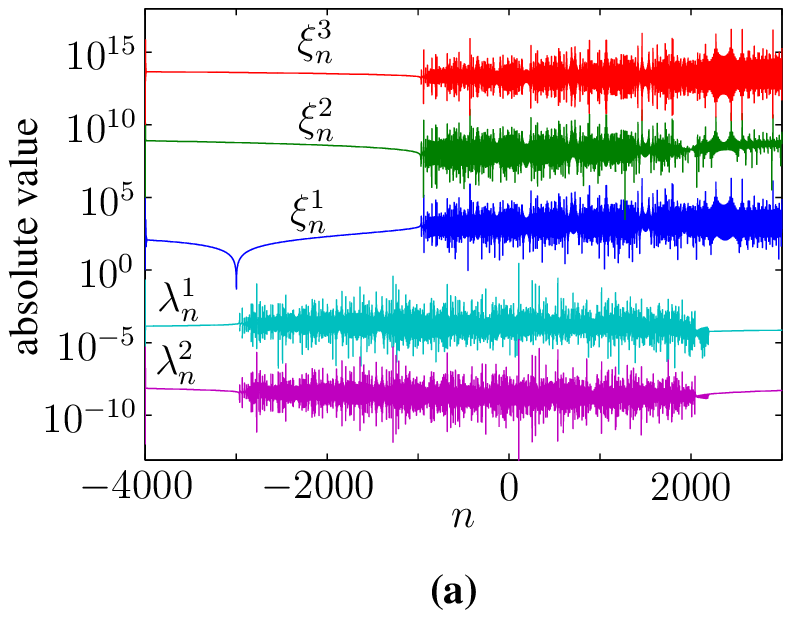}\hspace{1cm}
\includegraphics[width=0.45\linewidth]{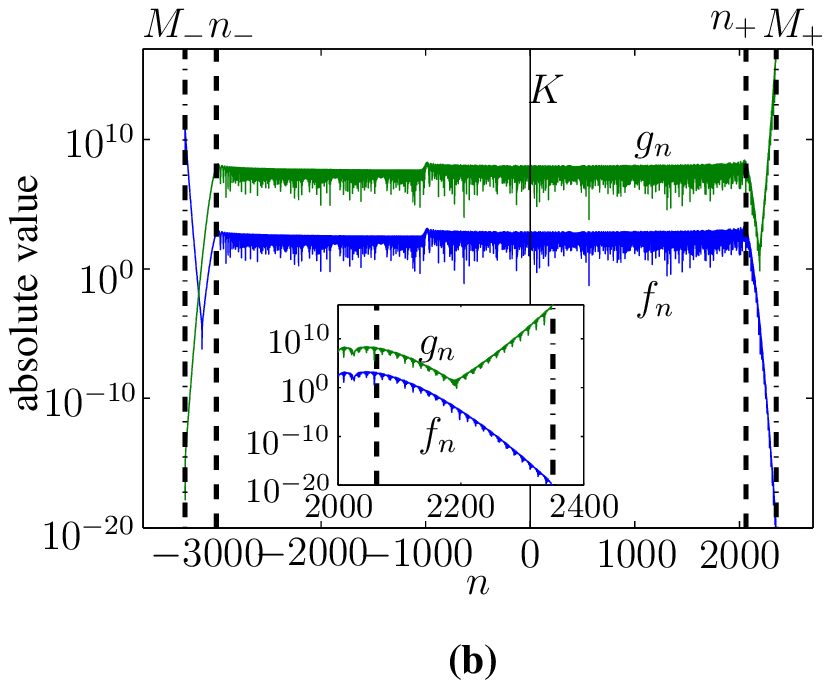}
\\[2ex]
\caption{\label{fig3}
(Color online.) Panel (a) shows the absolute value of the coefficients from
Eq.~\eqref{zeta_eta_xi_backwards} and Eq.~\eqref{kappa_lambda_backwards}, used
for the transformed recurrence relations \eqref{four-term_backwards} and
\eqref{three-term_backwards}. Here,
a starting index of $M_-=-4000$ and initial
values of $\xi_{M_-}^{1,2,3}=\lambda_{M_-}^{1,2}=1$ were used. Note that for
better visibility, all curves except $\xi_n^1$ have been shifted vertically 
on the logarithmic ordinate axis by
multiplication with a suitable constant ($10^{10}$ for $\xi_n^3$, $10^5$ for
$\xi_n^2$, $10^{-7}$ for $\lambda_n^1$, and $10^{-12}$ for $\lambda_n^2$). In
panel (b), we display the absolute values of $f_n$ and $g_n$, which result
after completing step 3 in the algorithm described, i.e. before normalizing.
The dash-dotted lines indicate the starting indices $M_\pm$, here $M_-=-3300$,
and $M_+=2350$. The cutoff indices $n_-=-3000$, $n_+=2063$ are plotted with
dashed lines.  An example of a suitable ``matching index'' $K=0$ (see step 4 of
the algorithm) is drawn by a solid line. The initial values used to calculate
the curves were $f_{M_+}=10^{-20}/2$, $f_{M_++1}=0$, $f_{M_++2}=10^{-20}$, and
$g_{M_+}=0$, $g_{M_++1}=10^{17}$. Note that in panel (b), no vertical
shifting was applied.  The inset shows a magnification of the cutoff region for
positive $n$, where the diverging behavior of $g_n$ for $n>n_+$ is clearly
seen. In both graphs we have $\argx=\argy=10^3$, same as in Fig.~\ref{fig2}.} 
\end{center}
\end{figure*}
To illustrate some of the intermediate steps of the algorithm, we show in 
Fig.~\ref{fig3} the typical behavior of the coefficients $\xi_n^{1,2,3}$ and 
$\lambda_n^{1,2}$ calculated in step 2, and also the result after step 3, 
before normalization of the arrays $f_n$ and $g_n$. Concluding the description 
of our recursive algorithm, we summarize the different integer indices which 
occur in the problem, which is useful to have in mind in the ensuing 
discussion: 
$n_-$ and $n_+$ are the negative and positive cutoff indices, respectively, and 
are fixed by the values of the arguments $\argx$ and $\argy$ through 
Eq.~\eqref{nminus_nplus}. $M_-$ and $M_+$ are the negative and positive 
starting 
indices, respectively. For the algorithm to converge, they should be chosen 
such that $M_-<n_-$, and $M_+>n_+$. The accuracy of the computed approximation 
to $\gbnxy$ will increase if the distances $n_--M_-$, $M_+-n_+$ are increased 
(see Sec.~\ref{Results_and_discussion}). $K$ is a matching index, where the 
solutions $f_n$ and $g_n$ computed with different recurrence relations should 
be 
matched, and should satisfy $n_-<K<n_+$. Finally, $\nmin$ and $\nmax$ are the 
indices between which numerical values for $\gbnxy$ are sought. Except for the 
requirements $\nmin>M_-$, $\nmax<M_+$, and $\nmin\le\nmax$, they can be 
arbitrarily 
chosen. The usual requirement is however $\nmin\le n_-$, $\nmax\ge n_+$, which 
in that case implies 
the following inequality chain for the different indices involved:
\begin{equation}\label{inequality_chain}
M_-<\nmin\le n_-<K<n_+\le\nmax<M_+.
\end{equation}

%
\subsection{Demonstration of numerical stability}
\label{Demonstration_of_numerical_stability}
In this subsection we show, using arguments similar
to those in \cite{Oliv1968}, that the previously presented algorithm 
is numerically stable.
Since the functions $\gbn{n}$, $Y_n$, $X_n$ and $Z_n$ (see
Fig.~\ref{fig2}) form a complete set of functions with respect
to the recurrence relation \eqref{recrel}, we
can decompose any solution $y_n$ to the four-term recurrence relation 
\eqref{four-term_backwards} as
\begin{equation}\label{decomposition}
y_n=a_1 \, \gbn{n} +a_2 \, Y_n +a_3 \, X_n +a_4 \, Z_n.
\end{equation}
The constants $a_1$, $a_2$, $a_3$, $a_4$ 
can be found from the initial conditions.
For general $i$ in the range $N-2 \le i\le N$, where $N$ is a general starting 
index (later we will take $N=M_-$), we have
\begin{equation}
y_i =a_1 \, \gbn{i} +a_2 \, Y_i +a_3 \, X_i +a_4 \, Z_i,
\end{equation}
but we can rewrite $y_{N+1}$ using the four-term recurrence 
in Eq.~\eqref{four-term_backwards} as
\begin{equation}
\label{initial_conditions_proof}
\begin{split}
& y_{N+1} = - \frac{1}{2\argy} \,
\left( \xi^1_{N} \, y_{N}+\xi^2_{N} \, y_{N-1}+\xi^3_{N} \, y_{N-2}
\right)
\\[2ex]
& \quad = a_1 \, \gbn{N+1} + a_2 \, Y_{N+1} + a_3 \, X_{N+1} +
a_4 \, Z_{N+1},
\end{split}
\end{equation}
for fixed starting integer $N$. 

If we now for simplicity take 
the initial value at the upper boundary of the recursion $y_{N+1}=0$, by
selecting the initial values 
$\xi^{1,2,3}_N$ for the coefficients accordingly,
then we can choose (provided the system \eqref{initial_conditions_proof} is
nonsingular, so that a solution exists) three sets of initial conditions
$y^t_i$, $1\le t\le3$, $N-2 \le i\le N$, so that depending on which set is
chosen, the constants $a_j$ in Eq.~\eqref{decomposition} are
\begin{equation}
\label{requir}
a_j=\delta_{jt},\qquad 
1\le j\le 3, \quad
1\le t\le 3,
\end{equation}
where $\delta_{jt}$ is the Kronecker delta, leading to the 
solutions $y^t_n$ with $1\le t\le 3$.
By requiring \eqref{requir}, we have implicitly reduced the 
solution to a linear combination of just two 
solutions, with nonvanishing components of one of $J_n$, $X_n$, $Y_n$
on the one hand, and $Z_n$ on the other hand.
The remaining constant $a_4$ is obtained, for each set, from
\begin{equation}
y^t_{N+1}=0=\delta_{1t} \gbn{N+1} +\delta_{2t} Y_{N+1} +\delta_{3t} X_{N+1}
+a_4Z_{N+1} \,,
\end{equation}
assuming $Z_{N+1}\neq 0$.
Because we have reduced the solutions $y^t_n$ to be linear 
combinations of just two functions, we immediately see that  
the three sets of initial values correspond to the three fundamental solutions
$y_n^{1,2,3}$ to the four-term recurrence relation
\eqref{four-term_backwards},
\begin{equation}
\begin{split}
y^1_n&=\gbn{n}-\frac{\gbn{N+1}}{Z_{N+1}}Z_n,\\
y^2_n&=Y_n-\frac{Y_{N+1}}{Z_{N+1}}Z_n,\\
y^3_n&=X_n-\frac{X_{N+1}}{Z_{N+1}}Z_n.
\end{split}
\end{equation}
If now $N$ is taken small enough, $N=M_-<n_-$, by virtue of
Eq.~\eqref{relative_amplitudes}, the three fundamental solutions $y_n^{1,2,3}$
turn to the three functions $\gbn{n}$, $Y_n$, and $X_n$. 
We have basically eliminated the unwanted solution $Z_n$
by rewriting the five-term recurrence \eqref{recrel} into a 
four-term recurrence relation \eqref{four-term_backwards}.

In other words, the
reduced four-term recurrence relation \eqref{four-term_backwards}, with the
coefficients $\xi^{1,2,3}_n$ evaluated according to
\eqref{zeta_eta_xi_backwards} in the direction of increasing $n$ from initial
values $\xi^{1,2,3}_{M_-}$, $M_-<n_-$ with a safety margin, 
has to a very good approximation the three
functions $\gbn{n}$, $Y_n$, and $X_n$ as fundamental solutions. This means that
a solution $f_n$ to the recurrence relation \eqref{four-term_backwards}, started
with initial values $f_l$, $M_+\le l\le M_++2$, $M_+>n_+$ with a safety margin, 
and applied in the
direction of decreasing $n$ will be almost proportional to $\gbn{n}$ for
$n<M_+$, by the same arguments as in subsection
\ref{Millers_algorithm_for_the_usual_Bessel_function},
because after having eliminated $Z_n$, the wanted solution 
$J_n$ is the only one which is suppressed for $n \to \infty$.
This is however only true down
to the negative cutoff index $n_-$ below which 
an admixture of the other unwanted solution $Y_n$ 
takes over, see Fig.~\ref{fig3}.

Similarly, for the three-term recurrence relation \eqref{three-term_backwards},
we can write a generic solution $v_n$ in terms of the three fundamental
solutions to the four-term recurrence relation \eqref{four-term_backwards},
\begin{equation}
v_n=b_1 \, \gbn{n}+ b_2 \, X_n + b_3 \, Y_n.
\end{equation}
Again, there exist two sets $v^s_j$, $1\le s\le 2$, $N-1\le j\le N+1$, of
initial conditions, with $v^{1,2}_{N+1}=0$, so that 
\begin{equation}
b_j=\delta_{sj}.
\end{equation}
The two fundamental solutions to Eq.~\eqref{four-term_backwards} are therefore
\begin{equation}
\begin{split}
v^1_n & = \gbn{n}-\frac{\gbn{N+1}}{Y_{N+1}}\,Y_n,\\
v^2_n & = X_n-\frac{X_{N+1}}{Y_{N+1}} \, Y_n \,.
\end{split}
\end{equation}
Thus, provided the recurrence for the coefficients 
of the three-term recurrence given in 
Eq.~\eqref{kappa_lambda_backwards} is started at sufficiently small,
negative $N=M_-<n_-$, and
applied in the forward direction, a solution $g_n$ to the three-term recurrence
relation \eqref{three-term_backwards}, started at a large $M_+>n_+$ and
performed in the direction of decreasing $n$,
will, to a good approximation, be
proportional to $\gbn{n}$ for $n<n_-$. Combining the solution $f_n$ to the
four-term equation \eqref{four-term_backwards} with the solution $g_n$ to the
three-term equation \eqref{three-term_backwards} 
at the matching index $K$, where $n_-<K<n_+$ then yields a solution
proportional to $\gbn{n}$ for all $n$, $\nmin\le n\le \nmax$. The
proportionality constant is found using the sum rule \eqref{sumruleJnxy}.

Having settled the question of convergence, we comment briefly on how to
numerically calculate the complementary solutions $Y_n$, $X_n$, and $Z_n$, shown
in Fig.~\ref{fig2}. We assume the most interesting case 1,
$\argx<8\argy$. The function $X_n$ can be computed by using the original
recurrence relation \eqref{recrel} in the direction of increasing
$n$, starting at an index $N<-2\argy +\argx$, i.e. in region $b_1$. Here $X_n$ 
quickly outgrows the
other solutions to leave only the ``pure'' $X_n$ after a few iterations. For
$Z_n$, we similarly use the original recurrence relation
\eqref{recrel}, but this time in the direction of decreasing $n$,
and starting at a large positive index $N>n_+$ 
[for the definition of $n_+$, see Eq.~\eqref{nminus_nplus}]. 
However, in this region $\gbn{n}$ grows as fast as
$Z_n$, and a solution $y_n$ calculated this way will be a linear combination
$y_n=a_1 \, \gbn{n}+a_2\,Z_n$ for $n>-2\argy+\argx$, the constants $a_{1,2}$ 
depending on the
initial values. For $n<-2\argy+\argx$ (in region $b_1$), $Z_n$ grows faster 
with 
decreasing $n$ than the other
fundamental solutions, so that here  $y_n=a_2\,Z_n$. 
Finally, using the four-term
relation \eqref{four-term_backwards} in the  backward direction, starting at
index $N<n_+$ in region $c_1$, yields a solution $x_n=a_1\gbn{n} +a_2Y_n+a_3 X_n$ 
for $-2\argy
+\argx <n<n_+$, $x_n =a_1\gbn{n} +a_2Y_n$ for $n_-<n<-2\argy +\argx$ and
$x_n=a_2Y_n$ for $n<n_-$.

%
%
\section{Discussion}
\label{Results_and_discussion}
%
%
\subsection{Accuracy}

\begin{figure}
\begin{center}
\includegraphics*[width=0.45\linewidth]{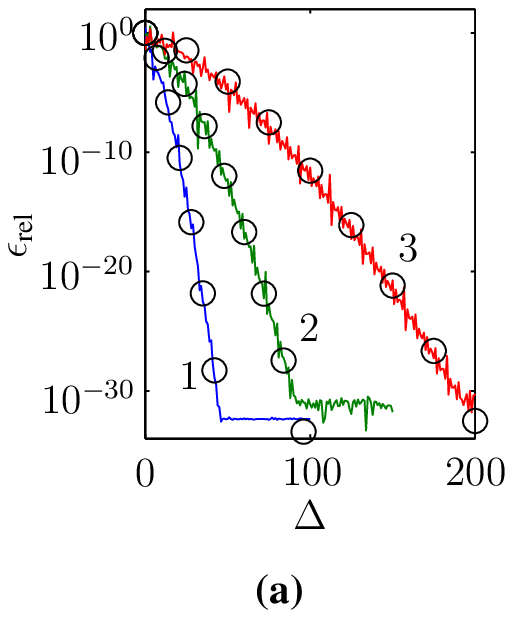}
%
\includegraphics*[width=0.45\linewidth]{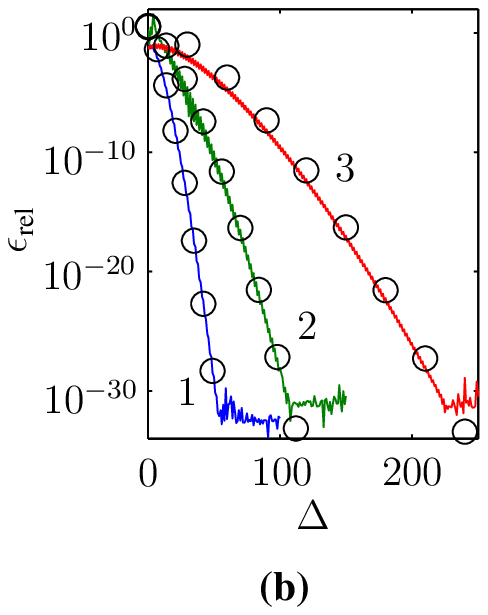}
\\[2ex]
\includegraphics*[width=0.45\linewidth]{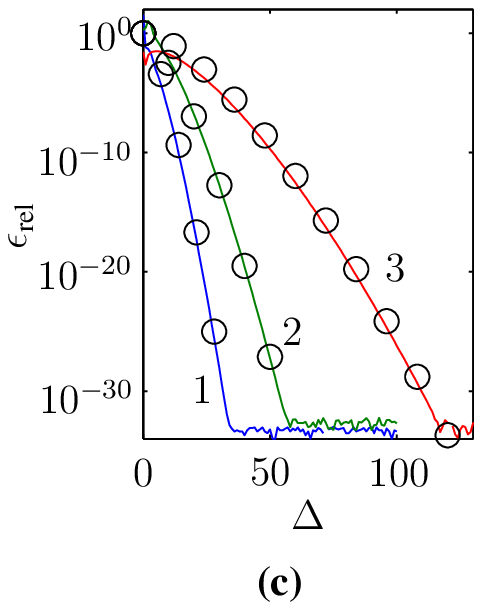} 
%
\includegraphics*[width=0.45\linewidth]{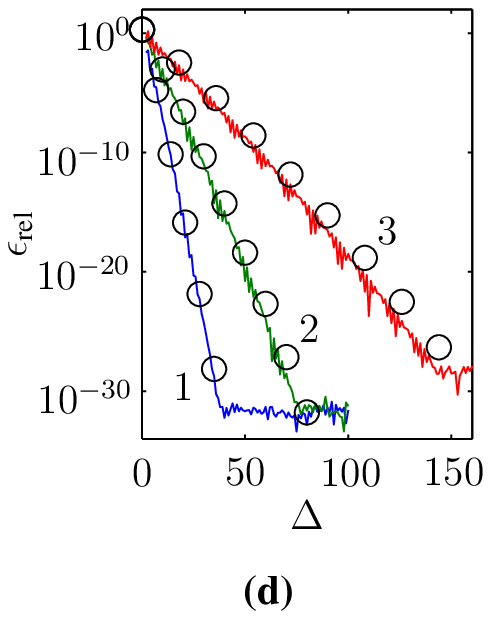}
\caption{\label{fig4}
(Color online.) 
The relative error $\epsilon_{\textrm{rel}}$,
as defined in Eq.~\eqref{relative_error}, 
as a function of the ``safety margin'' parameter
$\Delta$ [see Eq.~\eqref{definition_Delta}]. 
In each of the different parameter ranges considered, 
an exponential decrease of the obtained error with the 
safety margin parameter is observed, 
demonstrating the applicability of the 
recursive method. In (a), we consider parameters such that $x=2y$, in 
(b) we have large ratio $y/x$, in (c) small ratio $y/x$, and
in (d) results for the cutoff region are presented. Detailed explanation 
of the parameter regions considered is in
the text. In all graphs, black circles represent the approximation for 
the relative error obtained from Eqs.~\eqref{approximative_relative_error} and 
\eqref{approx_J_MoverY_M}.}
\end{center}
\end{figure}

It is necessary to investigate how the accuracy of the computed approximation
$h_n$ depends on the starting indices $M_-$, $M_+$. To this end, we define the
positive ``safety margin'' parameter $\Delta$ through
\begin{equation}
\label{definition_Delta}
M_-=\nmin-\Delta,\qquad M_+=\nmax +\Delta,
\end{equation}
so that specifying $\Delta$ fixes both the upper and the lower starting index,
and we also define the relative error
\begin{equation}
\label{relative_error}
\epsilon_{\textrm{rel}}=\left\lvert\frac{h_n-\gbnxy}{\gbnxy}\right\rvert.
\end{equation}
In Fig.~\ref{fig4}, we show the relative accuracy that can be obtained by the
method presented in this paper, as a function of $\Delta$, for different values
of the arguments $\argx$ and $\argy$, and different index $n$ in the obtained
array $h_n$.  We have numerically verified 
that a performance, similar to the one presented
in Fig.~\ref{fig4}, can be expected even close to zeros of
$\gbnxy$ (that is, for a general index $n$, $\nmin\le n\le\nmax$, where
$\gbnxy=0$ or $\gbnxy \approx 0$), 
although in this case the estimates remain valid only for the
absolute instead of the relative error.  Specifically, in the panels
(a)--(c) of Fig.~\ref{fig4}, we evaluate the relative error
$\epsilon_{\textrm{rel}}$ at $n=0$ and take $\nmin=n_-$, $\nmax=n_+$ [see
Eq.~\eqref{nminus_nplus}, and also the discussion preceding
Eq.~\eqref{inequality_chain}], which means that the recurrence is started at a
distance $\Delta$ from the cutoff indices.  The different curves in the graphs
correspond to the following values of $\argx$ and $\argy$: In (a), we have
$2\argy=\argx=10$ for curve 1 (blue line), $2\argy=\argx=10^2$ for curve 2
(green line), and $2\argy=\argx=10^3$ for curve 3 (red line). For these values
of $\argx$, $\argy$, the index $n=0$ corresponds to the border between the two
saddle point regions $b_1$ and $c_1$. We note that $\gbnxy$ cannot be
accurately evaluated in such border regions using the simple saddle point
approximation \cite{Leu1981,KoKlWi2006}, but that our method works well here.
In (b) we have $\argy=10\argx=10$ for curve 1 (blue line),
$\argy=10^2\argx=10^2$ for curve 2 (green line), and $\argy=10^3\argx=10^3$ for
curve 3 (red line), demonstrating the method for cases where the ratio
$\argy/\argx$ is large. In (c), we have instead a small ratio $\argy/\argx$:
$\argx=10\argy=10$ for curve 1 (blue line), $\argx=10^2\argy=10^2$ for curve 2
(green line), and $\argx=10^3\argy=10^3$ for curve 3 (red line). Finally, in
(d) we show the case where $\gbnxy$ is evaluated in the cutoff region, where
for all three curves $|\gbnxy|$ is of order $10^{-10}$. Here, we have
$\argy=\argx=10$, $n=\nmax=55$, $\nmin=n_--n+n_+=-64$ for curve 1 (blue line),
$\argy=\argx=10^2$, $n=\nmax=270$, $\nmin=n_--n+n_+=-364$ for curve 2 (green
line), and $\argy=\argx=10^3$, $n=\nmax=2200$, $\nmin=n_--n+n_+=-3137$ for
curve 3 (red line). The value $\nmin$ has in all cases in graph (d) been
chosen so that the distance $n_--\nmin$ equals $\nmax-n_+$. Recall that the
starting indices $M_\pm$ follows by fixing $\nmin$, $\nmax$, and $\Delta$, by
Eq.~\eqref{definition_Delta}. The black circles in the graphs (a)--(d) have
been obtained from Eq.~\eqref{approximative_relative_error}, using
approximation \eqref{approx_J_MoverY_M}. For the calculations, computer
arithmetic with 32 decimals was used.

%
%

An analytic formula for the relative error can be obtained by assuming that
after normalization, the calculated value $h_n$ is of the form (writing out
the dependence of $Y_n$ on the arguments $\argx$ and $\argy$ explicitly),
\begin{equation}\label{Ansatz_error}
h_n =
\gbnxy - \frac{\gb{M_-}{\argx}{\argy}}{Y_{M_-}(\argx,\argy)} \,
Y_n(\argx,\argy),
\end{equation}
for starting index $M_-<n_-$, similarly to the case for the ordinary Bessel
function [see Eq.~\eqref{solution_Miller_usual_Bessel_function}]. This is a
simplified assumption, since the total error in general is more complicated, but
Eq.~\eqref{Ansatz_error} can nevertheless be used to make practical predictions
about the dependence of $\epsilon_{\textrm{rel}}$ on $\Delta$. Equation
\eqref{Ansatz_error} yields for the approximative relative error
\begin{equation}\label{approximative_relative_error}
\epsilon_{\textrm{rel, app}} =
\left\lvert\frac{h_n-\gbnxy}{\gbnxy}\right\rvert =
\left\lvert\frac{\gb{M_-}{\argx}{\argy}}{Y_{M_-}
(\argx,\argy)} \,
\frac{Y_n(\argx,\argy)}{\gbnxy}\right\rvert.
\end{equation}
An approximation for the amplitude of $Y_{M_-}(\argx,\argy)$ for $M_-<n_-$ 
can be obtained from the
saddle point expression \eqref{asymptotic_approximation} for $\gbnxy$, but
reversing the sign of the real part of the argument of the exponential. If we
write the saddle point approximation of $Y_n(\argx,\argy)$ as
$Y_n(\argx,\argy)=G_+(n)+G_-(n)$, we have
\begin{equation}\label{approx_J_MoverY_M}
\begin{split}
& \left\lvert\frac{\gb{M_-}{\argx}{\argy}}{Y_{M_-}(\argx,\argy)}\right
\rvert \approx
\left\lvert\frac{F_+(M_-)+F_-(M_-)}{G_+(M_-)+G_-(M_-)}\right\rvert 
\approx 
\left\lvert\gb{M_-}{\argx}{\argy}\right\rvert^2\\
& \quad \approx \left(\ee^{-\left|\mathrm{Re}[\ii f(\theta_+)-
\ii M_-\theta_+]\right|}+
\ee^{-\left|\mathrm{Re}[\ii f(\theta_-)- \ii\, M_-\theta_-]\right|}\right)^2,
\end{split}
\end{equation}
where $f(\theta)$ is defined as 
after Eq.~\eqref{asymptotic_approximation}, and
$\theta_\pm$ denote the two different saddle point solutions from
Eq.~\eqref{saddle_point_equation}, with $n=M_-$. The last approximation in
Eq.~\eqref{approx_J_MoverY_M} neglects the preexponential factor and the
oscillating factor in the saddle point approximation
\eqref{asymptotic_approximation}, which is sufficient for an order-of-magnitude
estimate. The ratio ${Y_n(\argx,\argy)}/{\gbnxy}$ in
\eqref{approximative_relative_error} can be approximated with unity for $n$
in the oscillating region [graph (a), (b), and (c) in Fig.~\ref{fig4}],
and with the simplified saddle point approximation 
\eqref{approx_J_MoverY_M} for $n$ in the cutoff region [graph (d) in
Fig.~\ref{fig4}]. The approximation
\eqref{approximative_relative_error} together with
\eqref{approx_J_MoverY_M} for the relative error is plotted as circles in
Fig.~\ref{fig4}. Clearly the approximate formula can be used for practical
estimates of how far out the recurrence should be started
if a specific accuracy is sought for the array of generalized 
Bessel functions to be computed. Formula~\eqref{approx_J_MoverY_M} also
explains the exponential decrease in relative error observed in
Fig.~\ref{fig4}.

%
%
\subsection{Comparison with other methods}

Here we briefly comment on the performance of the presented algorithm as
compared to other ways of numerically evaluating the generalized Bessel
function. Let us compare to an alternative algorithm based on the 
evaluation of ordinary Bessel functions using the techniques 
outlined in Sec.~\ref{Millers_algorithm_for_the_usual_Bessel_function},
where we first calculate two arrays
$J_{2s+n}(\argx)$, $J_s(\argy)$ of ordinary Bessel functions by
Miller's algorithm and later calculate the generalized Bessel functions
using Eq.~\eqref{gen_bes_as_sum_usual_bes}.
Calculation of the arrays of ordinary Bessel functions
then requires two recurrence
runs, and to obtain the numerical value $\gbnxy$, in addition the sum
$\sum_{s=-\infty}^\infty J_{2s+n}(\argx)J_s(\argy)$ 
has to be performed. This means that since the
generalized Miller's algorithm requires two recurrence runs only, for
calculation of a {\it single} value $\gb{n_0}{\argx}{\argy}$, the two methods
demand a comparable amount of time. 
However, the calculation of a single 
generalized Bessel function is not the aim of our 
considerations: 
for the whole array $\gbnxy$, $\nmin\le n\le\nmax$, 
the reduction in computer time due to the elimination of the 
calculation of the sums $\sum_{s=-\infty}^\infty J_{2s+n}(\argx)J_s(\argy)$ 
leads to an order-of-magnitude gain with respect to 
computational resources while
the accuracy obtained by the two different methods is similar.

The second method with which to compare is the asymptotic expansion by
integration through the saddle points, as presented in \cite{Leu1981}. For
evaluation of a single value $\gb{n_0}{\argx}{\argy}$, with moderate
accuracy demands, the saddle-point integration is of course the best method,
especially for large values of the parameters $n$, $\argx$ and $\argy$. The
drawback of this method is the relatively 
complex implementation \cite{Leu1981},
and in addition, an increase in the accuracy of a saddle-point method 
typically is a nontrivial task  which involves higher-order
expansions of the integrand about the saddle point,
and this typically leads to very complicated
analytic expressions for higher orders, especially for an 
integrand with a nontrivial structure as in Eq.~\eqref{def_Jnxy}.
See however \cite{HuVa2006} for a possibly simpler numerical method,
the ``numerical steepest descent method''.
In any case, if the complete array
$\gbnxy$, $\nmin\le n\le\nmax$ is sought to high accuracy,
as it is the case for second-order laser-related problems, 
then our method is necessarily better, since
the time spent on one recursive step is very brief.  

%
%
\section{Illustrative Considerations for the Dirac--Volkov Solutions}
\label{Illustrative_Considerations_for_the_Dirac--Volkov_Solutions}

In this Section, we consider the Volkov solution, the analytic solution to the
Dirac (or Klein-Gordon) equation coupled to an external, plane-wave laser
field. We show that the generalized Bessel functions can be directly
interpreted as the amplitudes for discrete energy levels of a quantum
laser-dressed electron,  corresponding to the absorption or
emission of a specific number of laser photons.

%
%
\subsection{Physical origin of the recurrence relation}
\label{Physical_origin_of_the_recrel}

There is a direct, physical way to derive the recurrence relation satisfied by
the generalized Bessel function, in the context of relativistic laser-matter
interactions. The result of this approach defines $\gbnxy$ in terms of the
recurrence relation and a normalization condition, even on the level of
spinless particles, i.e.~on the level of Klein-Gordon theory.  In this section,
we set $\hbar=c=1$, denote the electron's charge and mass by $e=-|e|$ and $m$,
respectively, and write dot products between relativistic four-vectors as
$u\cdot v=u_\mu v^\mu=u^0 v^0-\vect{u}\cdot \vect{v}$, for two four-vectors
$u^\mu$ and $v^\mu$. The space-time coordinate is denoted by $z^\mu = (t, \vec
x)$, in order not to cause confusion with the argument $\argx $ of $\gbnxy$,
and $k \cdot z = \omega\,t - \vec k \cdot \vec x$ is the phase of the laser
field. The $4\times4$ Dirac gamma matrices are written as $\gamma^\mu$.

Let us consider the Klein-Gordon equation $\left[\left( \ii \partial_z - e A
\right)^2 -m^2\right]\psi(z) = 0$ for the interaction of a spinless particle
of charge $e$
with an external laser field of linear polarization $A^\mu(z)=a^\mu \cos
(k\cdot z)$,
\begin{equation}
\begin{split}
\label{Klein-Gordon_Floquet_ansatz}\
& \bigg[-\partial^2_z -
2 \ii e \cos (k\cdot z) \, a \cdot \partial_z \\
& \quad 
+ \frac{e^2 \,a^2}{2}\cos (2k\cdot z)
-m^2-\frac{|e^2a^2|}{2}\bigg]\psi(z) = 0 \,.
\end{split}
\end{equation}
Here $a^\mu=(0,\vect{a})$ is the polarization vector, and $k^\mu=(\omega,\vec
k)$ is the propagation wave vector of the laser field, with $k\cdot k=k\cdot
a=0$.  We also introduce the four-vector $q^\mu$, the so-called effective
momentum \cite{LaLi1982}, which fulfills
\begin{equation}
q^2=m^2+ \tfrac{1}{2} \, |e^2a^2| \,.
\end{equation}
We now insert the Floquet ansatz \cite{ChuTel2004} for the wave function
\begin{equation}\label{Floquet_ansatz}
\psi(z) = \ee^{-\ii q\cdot z} \sum_s B_s \, \ee^{-\ii s k\cdot z} \,,
\end{equation}
where the coefficients $B_n$ are independent of $z^\mu$, into 
Eq.~\eqref{Klein-Gordon_Floquet_ansatz}.
From this representation, we see that the factor $\ee^{-\ii s k\cdot z}$,
actually has the same form as a phase factor characterizing the
absorption of $s$ laser photons from the laser field,
as we integrate over the Minkowski coordinate $z$ in the 
calculation of an $S$-matrix element.
For negative $s$, we instead have emission into the laser mode.
Equation \eqref{Floquet_ansatz} also 
leads to a relation for the coefficients $B_s$,
\begin{equation}
\begin{split}
\label{a_relation_for_the_coefficients_A_s}
& \sum_{s = -\infty}^{\infty}
\left[\argx \cos(\phi) - 2\argy \cos(2\phi)-s\right] \,
B_s \, \ee^{-\ii s\phi} = 0\,,
\\
& \argx = \frac{e \, a\cdot q}{k\cdot q}\,, \qquad
\argy = \frac{e^2\,a^2}{8k\cdot q}\,, \qquad
\phi = k \cdot z \,.
\end{split}
\end{equation}
Multiplying Eq.~\eqref{a_relation_for_the_coefficients_A_s} with
$\ee^{\ii n \phi}$, and integrating over one period, we obtain the recurrence
relation \eqref{recrel}, if we identify 
\begin{equation}
B_n \equiv \gbnxy \,.
\end{equation}
For the wave function~\eqref{Floquet_ansatz} constructed from the solution to
the recurrence relation \eqref{recrel} to be finite, we  must demand $\gbnxy$
to be normalizable. This is expressed by the condition \eqref{sumruleJnxy}.
Furthermore, using the property \eqref{fourier} of $\gbnxy$, we can perform the
sum over $s$ in \eqref{Floquet_ansatz}, with the result
\begin{equation}
\psi(z)=\ee^{-\ii q\cdot z- \ii x\sin\phi+ \ii y\sin(2\phi)},
\end{equation}
which is the form in which the Volkov solution is 
usually presented \cite{LaLi1982}.

For comparison, the solution $\Psi(z)$ to the Dirac equation in presence of a
linearly polarized laser field,
\begin{equation}
\left(\ii \, \gamma\cdot\partial_z - 
e\, a\cdot\gamma \, \cos\phi -m \right) \, \Psi(z)=0
\end{equation}
reads
\begin{equation}
\begin{split}
\Psi(z) = 
\ee^{-\ii q\cdot z} \sum_{s=-\infty}^\infty
\Big(&\frac{ek\cdot\gamma \, a\cdot \gamma}{4 \, k\cdot q} \,
[\gb{s+1}{\argx}{\argy}
+\gb{s-1}{\argx}{\argy}]\\
&+\gb{s}{\argx}{\argy}\Big) \, 
\ee^{-\ii \, s \, \phi} \, u_q \,, 
\end{split}
\end{equation}
where $\argx$, $\argy$ are defined in 
Eq.~\eqref{a_relation_for_the_coefficients_A_s}, 
and $u_q$ is a Dirac bispinor satisfying
\begin{equation}
\left(\gamma\cdot q+2 \, \argy \, \gamma\cdot k-m\right) \, u_q = 0 \,.
\end{equation}
The four-vector $p^\mu=q^\mu+2\argy k^\mu$ can be identified as the asymptotic
momentum of the particle, or the residual momentum as the laser field is turned
off.

%
%
\begin{figure}
\begin{center}
\includegraphics*[width=0.95\linewidth]{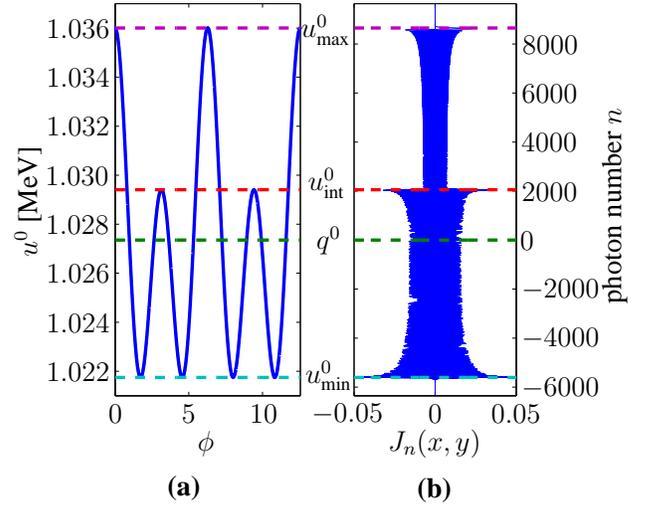}
\caption{\label{fig5}
(Color online.) Illustration of the classical-quantum correspondence of a
laser-dressed electron.  Shown to the left [panel (a)] 
with a solid blue line is the
classical energy $u^0$ [see Eq.~\eqref{classical_kinetic_momentum}] as a
function of the phase $\phi$, for $\omega=1$ eV, laser intensity $I=10^{16}$
W/cm$^2$ (corresponding to $|e \, a|/m=0.1$ where $a$ is the 
laser polarization four-vector), 
initial energy $p^0=2m$, and initial
angle $\theta=0.54^\circ$, with $\vec p\cdot \vec k=|\vec
p|\,\omega\cos\theta$. These parameters give $\argx=3.3\times 10^3$, and
$\argy=-2.7\times 10^3$ for the arguments of $\gbnxy$. The dashed lines show
the minimum classical energy $u^0_{\textrm{min}}$, the maximum classical energy
$u^0_{\textrm{max}}$, the average energy $q^0$, and the intermediate energy
level (local maximum) $u^0_{\textrm{int}}$, as indicated in the center of the
figure.  In panel (b), we display the quantum mechanical amplitude $\gbnxy$ of
energy level $n$, which has energy $q^0+n\omega$ [see
Eq.~\eqref{Floquet_ansatz}]. Here positive index $n$ corresponds to absorbing
$n$ number of photons from the laser field, while negative $n$ means emitting
$|n|$ number of photons into the laser mode. The graph is arranged such that
$n=0$ corresponds to the average energy $q^0$. The cutoff indices are nicely
reproduced by the classical maxima and minima. } 
\end{center}
\end{figure}

%
%
\subsection{Classical-quantum correspondence of Volkov states}

It follows from the expression~\eqref{Floquet_ansatz}, that a quantum Volkov
state (we consider the spinless case for simplicity) can be regarded as a
superposition of an infinite number of plane waves with definite, discrete,
four-momenta $q^\mu+nk^\mu$.  The amplitude to find the particle with
four-momentum $q^\mu+nk^\mu$ is given by $\gbnxy$, with $\argx$ and $\argy$ as
in Eq.~\eqref{a_relation_for_the_coefficients_A_s}. Therefore, it might seem
that the particle can acquire arbitrarily high energy in the field. That this
is not so, follows from the exponential decay of $\gbnxy$ beyond the cutoff
indices, as discussed in subsection~\ref{Saddle_point_considerations}. In the
following, we show that the cutoff indices can also be derived as the lowest
and highest energy of a classical particle moving in a laser field. To this
end, we first recall the classical, relativistic equations of motion of a
particle of charge $e$ and mass $m$, moving in the laser potential $A^\mu=a^\mu
\cos \phi$:
\begin{equation}
\frac{\dd u^\mu}{\dd \tau} =
\frac{e}{m}(a^\mu \, k\cdot u - k^\mu \, a\cdot u)\sin\phi,
\end{equation}
where $u^\mu$ is the kinetic momentum, and $\tau$ is the proper time. The
solution reads \cite{SaSc1970,Me1971}, assuming initial phase $\phi_0=\pi/2$,
\begin{equation}\label{classical_kinetic_momentum}
u^\mu = p^\mu + x \, k^\mu \, \cos\phi - 
4 \, y \, k^\mu \, \cos^2\phi - e \, a^\mu \, \cos\phi \,,
\end{equation}
where $p^\mu$ is the asymptotic momentum. Note that as $u^\mu$ is the physical
momentum, it is gauge invariant under $A^\mu\to A^\mu+\Lambda k^\mu$, where
$\Lambda$ is an arbitrary function. The phase average is exactly the effective
momentum, $\overline{u^\mu}=p^\mu-k^\mu e^2a^2/(4k\cdot p)=q^\mu$.  In
Fig.~\ref{fig5}, we consider the energy $u^0$ as a function of the phase $\phi$
and compare it with the discrete energy levels $q^0 +
n \, \omega$ of the quantum
wave function. We see that the maximal and minimal energy of the classical
particle correspond exactly to the cutoff indices of the generalized Bessel
function. The probability for the quantum particle to have an energy larger (or
smaller) than the classically allowed energy is thus exponentially small.
Interestingly, the local maxima of the classical energy $u^0$, labeled
$u^0_{\textrm{int}}$ in Fig.~\ref{fig5}, coincide with the transition between
the two different saddle point regions of $\gbnxy$. 
 
%
%
\section{Conclusions}
\label{Conclusions_and_outlook}

We have presented a recursive algorithm for numerical evaluation of the
generalized Bessel function $\gbnxy$,
which is important for laser-physics related problems,
where the evaluation of large arrays of generalized
Bessel functions is crucial. In general, we can say that the 
laser parameters fix the arguments $\argx$ and $\argy$ of the 
generalized Bessel function $\gbnxy$, while the index $n$ 
characterizes the number of exchanged laser photons. 

As evident from Figs.~\ref{fig1} and~\ref{fig2}, complementary solutions $Y_n$,
$X_n$, and $Z_n$ to the recurrence relation \eqref{recrel} satisfied by
$\gbnxy$ are central to our algorithm.  By removing the sources of numerical
instability, which are the exponentially growing complementary solutions, in a
first recurrence run, we are able to construct a stable recursive algorithm,
similar to Miller's algorithm for the ordinary Bessel function, but suitably
enhanced for the generalized Bessel function. Numerical stability is
demonstrated, and the obtainable accuracy is studied numerically and by an
approximate formula (see Sec.~\ref{Results_and_discussion}).  The algorithm is
useful especially when a large number of generalized Bessel function of
different index, but of the same argument, is to be generated.
As is evident from the discussion in 
Sec.~\ref{Illustrative_Considerations_for_the_Dirac--Volkov_Solutions},
a fast and accurate calculation of generalized Bessel functions 
leads to a quantitative understanding of the quantum-classical
correspondence for a laser-dressed electron.

%
%
\section{Acknowledgment}

U.D.J.~acknowledges support by the Deutsche
Forsch\-ungs\-ge\-mein\-schaft (Heisenberg program)
during early stages of this work.
%
%

\appendix*
\section{Integral representation of the complementary solutions}

In this Appendix, we present the expressions for the integral
representations of the complementary solutions
$Y_n(\argx,\argy)$, $Z_n(\argx,\argy)$,
and $X_n(\argx,\argy)$ to the recurrence
relation~\eqref{recrel}, without giving any details
about the mathematical considerations 
which lead to these representations.  The
integrals read
\begin{align}
\label{Y_n_intrep}
Y_n(\argx,\argy)
=& -\frac{1}{\pi}\int_0^\infty 
[\cos(n\pi/2-\argx\cosh \theta )
\ee^{n\theta}
\nonumber\\
&\quad\qquad\quad+(-1)^n \ee^{-n\theta-\argx \sinh\theta}]
\ee^{-\argy \sinh 2\theta}\dd\theta
\nonumber\\
&\quad-\frac{1}{\pi}\int_{\pi/2}^\pi
 \sin({n\theta-\argx\sin\theta +\argy\sin 2\theta})\dd \theta,
\\
\label{X_n_intrep}
X_n(\argx,\argy) =& -\frac{1}{\pi}\int_0^\infty 
\sin(n\pi/2-\argx\cosh \theta)\,
\ee^{n\theta-\argy \sinh 2\theta}\dd\theta
\nonumber\\
&\quad-\frac{1}{\pi}\int_0^{\pi/2} \cos({n\theta-\argx\sin\theta
 +\argy\sin 2\theta})\dd \theta,
\\
\label{Z_n_intrep}
Z_n(\argx,\argy)
=& \; -\frac{1}{\pi}\int_{0}^\pi \sin({n\theta-\argx\sin\theta 
+\argy\sin 2\theta})\dd \theta
\nonumber\\
&\quad-\frac{1}{\pi}\int_0^\infty
 [(-1)^n\ee^{-\argx\sinh\theta}-\ee^{\argx\sinh\theta}]
\nonumber\\
&\qquad\qquad\quad\times \ee^{-n\theta-\argy\sinh2\theta}
 \dd\theta.
\end{align}

Recall that we consider non-zero, positive
values of the arguments $\argx$ and $\argy$, and an 
arbitrary  integer $n$.
By partial integration, the functions \eqref{Y_n_intrep}---\eqref{Z_n_intrep}
verify the recurrence relation~\eqref{recrel}. The prefactor has been
selected for each case so that the functions $X_n(\argx,\argy)$, 
$Y_n(\argx,\argy)$,
and $Z_n(\argx,\argy)$ have the same amplitude as $\gbnxy$ in 
the oscillating region, and this choice also implies that 
the functions $Y_n(\argx \to 0,\argy)$ and $Z_n(\argx \to 0,\argy)$,
for even and odd $n$, respectively, 
can be expressed as Neumann functions of fractional order.
(The latter statement is also given here without proof.)
A more detailed discussion of the mathematical properties
of the four functions defined by the integral representations
\eqref{def_Jnxy},~\eqref{Y_n_intrep},~\eqref{X_n_intrep} 
and~\eqref{Z_n_intrep} will be given elsewhere.
For all considerations reported in the current
article, the detailed knowledge of the integral representations
is not necessary; it is sufficient to know the recurrence
relation~\eqref{recrel} that they fulfill.

\end{document}